# Gaseous Tritium Activity Monitoring with Scintillators (GaTAMoS)

Florian Priester[a], Alexander Marsteller[a], Johanna Wydra[a], Robin Größle[a]

[a]*Karlsruhe Institute of Technology, Institute for Astroparticle Physics, Tritium Laboratory Karlsruhe, Hermann von-Helmholtz-Platz 1, 76344 Eggenstein-Leopoldshafen, Germany*

We present a novel device for the inline activity monitoring of gaseous tritium, named "*Ga*seous *T*ritium *A*ctivity *Mo*nitoring with *S*cintillators" (GaTAMOS). Unlike established methods that often rely on liquid scintillation or plastic scintillators with limited chemical resistance, this system utilizes a chemically inert, inorganic Terbium-activated Gadolinium Oxysulfide ($Gd_2O_2S$:Tb) ceramic. A specialized soldering technique was developed to join the ceramic scintillator to a stainless-steel flange, achieving a vacuum-tight seal (<$10^{-9}$ mbar·l/s) fully compatible with high-purity tritium applications. The device was characterized using pure tritium gas, demonstrating a linear response and high sensitivity across a dynamic range from <1 kBq to >10 GBq. A key advantage of the ceramic material over organic alternatives is its high tolerance to aggressive cleaning agents. We demonstrate that the inevitable tritium memory effect can be significantly mitigated by flushing the cell with ethanol, removing >95% of residual contamination within minutes. The proposed system offers a compact, robust, and cost-effective alternative to ionization chambers or BIXS systems for process monitoring in tritium handling facilities.

## 1. Introduction

Scintillation counting is a well-established method in various applications. Usually, a scintillating material is used to convert a high-energy primary particle (α, β or γ) to visible (secondary) photons, and the photon yield is proportional to the energy of the primary particle in most cases. The secondary photons are recorded with a light-sensitive detector like a photomultiplier tube (PMT), photo diode or similar, which is coupled to the scintillating material. The right choice and combination of scintillating material, detector and configuration can lead to very sensitive devices with single photon detection limits.

Common applications of scintillation counting include high energy physics detectors where mostly solid scintillators or fibres with a scintillating core are used [1][2], X-ray applications where a solid scintillator is coupled with a (digital) photo plate for technical, security or medical imaging [3], and environmental monitoring where a liquid scintillator is mixed with a sample [4].

In case of liquid scintillators, the measurement usually involves manual labour in an offline, low-throughput sample preparation process, and produces potentially contaminated, hazardous waste. This disadvantage is offset by providing the lowest possible level of detection for samples that can be dissolved in the liquid scintillator [4][5].

Solid scintillators widely used in particle detectors are in most cases made from an organic/plastic (e.g., polystyrene) matrix doped with an active component, like anthracene [6]. These materials can be made in large quantities for relatively low prices and machined easily into virtually any shape, allowing for large setups with arbitrary geometry. For low energy particles, the geometry, and ultimately, the surface exposed to the particle flux are limiting the lower level of detection. One downside of plastic scintillators, while not explicitly hygroscopic, is that they are not intended to be in direct contact with water for an extended period of time. Being an organic material, tolerance to acids, oils, solvents, aggressive gases as well as elevated temperatures is low [1][7].

This is where another group of solid scintillators comes in handy, mostly used for X-ray applications. Those consist of a ceramic matrix doped with an optical active element. One well-known group are terbium-activated GOS (Gadolinium oxysulfide, $Gd_2O_2S$:Tb) scintillators with peak emission wavelength in the visible (green) range, matching peak sensitivities of established photon detectors [8]. Those chemically highly stable scintillators can be machined to any shape, produced vacuum tight, coated using common thin film applications, and withstand most challenging conditions, like ultra-high vacuum, high flux of primary particles, the direct contact with radioactive gases/liquids, and all commonly used solvents (alcohol, isopropanol, acetone) and oil.

These promising properties make GOS scintillators an ideal candidate material for applications with tritiated and pure tritium gas streams.

## 2. Experimental description

The aim of the newly designed device is to address several problems arising when handling pure tritium or mixtures containing it:

a) General material compatibility, as tritium is a highly mobile gas species, causing damage to most polymeric materials.
b) Usually, a history-dependent memory effect results from the use of tritium. This means in particular, that the lower level of detection might shift towards higher pressure/activity levels during operation without appropriate countermeausures in place.
c) Gas mixture dependance. A gas-mixture-independent response is preffered, which means

*author's email: florian.priester@kit.edu*

less effort for calibration and a broader applicability of the method.

From this, a few general as well as some specialised requirements arise for the proposed device.

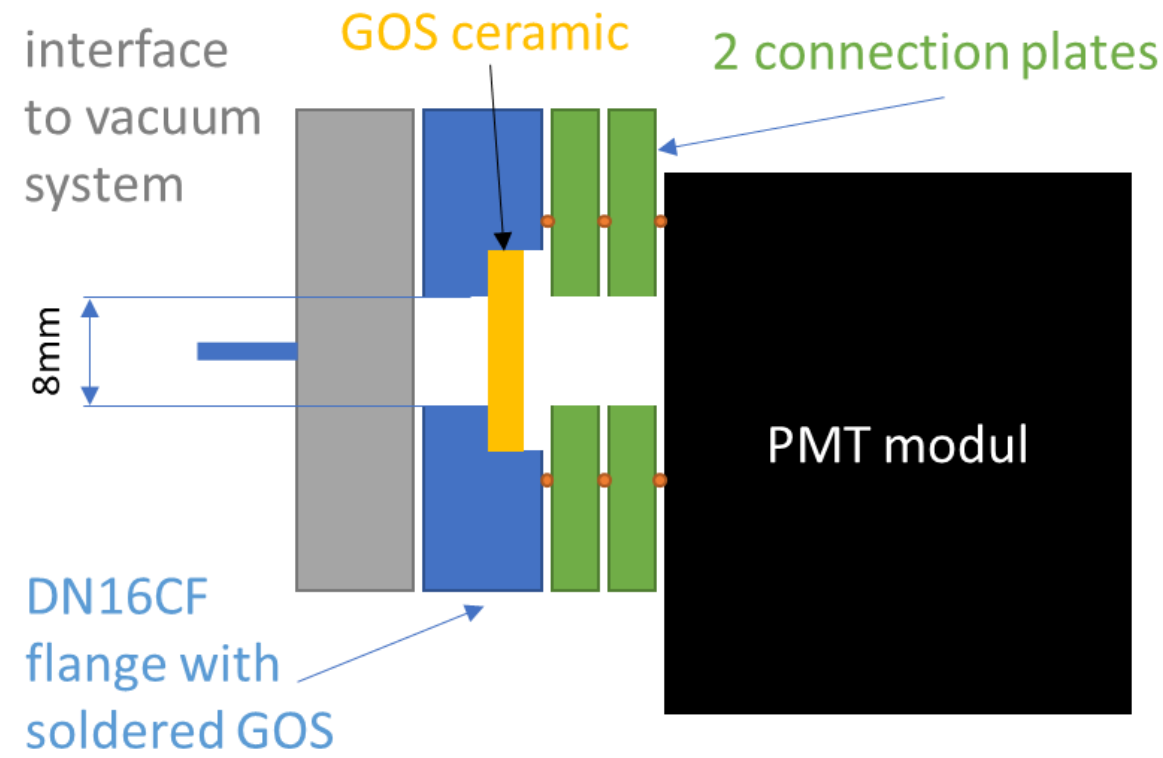

Figure 1: Sketch of a scintillation-based, tritium-compatible device.

### 2.1 Requirements

Handling tritium in significant amounts, such as hundreds of mbar of pure tritium gas, always brings the requirement of a very low single connection leak rate of $10^{-9}$ mbar·l/s (He equivalent). This can only be achieved using all metallic sealings.

Another important requirement is the material compatibility. In general, systems containing tritium must be made from metal, with glass and ceramic allowed as an exception. Systems not made entirely from metal must still be joined with a metallic seal.

To overcome possible memory effect issues, an easy possibility for decontamination should be possible. A high tolerance against liquid (ethanol, isopropanol, acetone) or gaseous (UV/ozone, air with high humidity) cleaning agents is a necessity.

Additional requirements arise from the use inside a glove box with limited space and the possibility of easy fitting into existing facilities, like a small footprint of the device, an easy installation, and the possibility to operate between vacuum ($10^{-6}$ mbar) up to ≈ 950 mbar in any orientation.

### 2.2 Technical details

As displayed in Figure 1, the proposed system consists of very few components.

As a detector, an-easy-to-operate Hamamatsu H11890 series photomultiplier tube was chosen. This device only needs a single USB connection to be operated. The internal electronics generate all necessary high voltage for operation and a photon counting circuit is integrated as well. The wavelength of peak sensitivity of this device matches well with the GOS scintillator emission, maximising efficiency. The dark count noise of 25-50 cps is low enough to not limit any application with significant amounts of tritium.

We are using a Ø 12 mm GOS:Tb disc with a thickness of 0.5 mm for all measurements presented in this publication. The chemically inert GOS ceramic fulfils all requirements regarding vacuum tightness and tritium compatibility but, to our knowledge, has not been tested with low-energy electrons. The scintillation yield is usually given in photons/MeV, with around 40,000-50,000 photons/MeV primary particle for our GOS. In the case of linear scaling, this translates to ≈ 250 photons/tritium decay with a mean β-energy of 5.7 keV. This is significantly above the dark noise equivalent of the PMT and therefore, easily detectable.

The main challenge to overcome is the tritium-compatible joint between the GOS ceramic and the vacuum flange (DN16CF) it is embedded to. We opted for the same technique developed earlier at TLK, which is used for joining fused silica to stainless steel. A total of three thin film layers (CrNi, FeNi, Au) are deposited on the stainless steel and on the GOS ceramic, while the GOS ceramic is masked with Ø 8 mm round opening to keep an active surface free from any deposit allowing tritium to directly interact with the GOS scintillator. The sputtering process was performed at company Siegert TFT (Germany).

GOS and stainless steel, both coated, are joined using a lead-containing soldering paste used for reflow soldering with a low melting point of ≈ 185°C. The soldering paste is applied manually, both parts slightly compressed and heated in an oven until thoroughly joined.

Detector and flange with scintillator are joined light-tight by a two-parted system allowing to remove the detector without opening the connection to the primary system. Since this connection is not primary-system relevant, standard rubber seals can be used to achieve a light tight system, which can be operated without an additional case. This allows for easy detector swap in case of an electrical failure or a replacement with another type of detector (e.g., photo diode), if desired.

To complete the setup, a gas independent pressure gauge (Wika WU20, 0…2 bar(a), uncertainty < 0.6% of full range) is attached for monitoring the filling of the cell with tritium.

Tritium supply is facilitated via another setup at TLK acting as an interface to the general tritium handling infrastructure.

## 3. Measurements

The device describe above was tested at TLK with pure tritium as described in the following.

### 3.1 Procedure

To start a series of measurements, tritium has to be transferred from the tritium transfer system to the setup. It is stored in a DN40CF vessel acting as an intermediate storage to become more independent.

From that buffer, which provides a pressure < 140 mbar, tritium is sent via a manual valve to the GaTAMoS setup. By this, it is possible to adjust discrete pressure settings or, by only slightly opening the valve ("leak-valve"), to achieve a continuous increase. Decreasing the pressure works in the same fashion with a tritium compatible scroll pump evacuating the combined volume of the intermediate storage vessel and the measurement cell.

The corresponding pressure values are logged by a dedicated data logger, while the PMT is operated, controlled, and the data recorded by a PC with Hamamatsu's software interface. Only the raw count rate is recorded with no corrections applied.

### 3.2 Response to tritium

Figure 2 shows the result of the first exposure of the GaTAMoS setup to tritium. Using pure (>98%) tritium, discrete pressure steps were adjusted. The graph shows the results of the very first measurement with tritium, thus the background at the beginning is at its lowest while the pressure reading of $10^{-3}$ mbar is an estimate from the ultimate pressure of the combined pumps used since the pressure sensor has a usable lower limit of ≈ 0.1 mbar. Even here, the outgassing of the interface, which has been operated with tritium previously, to the GaTAMoS setup was enough to elevate the initially determined count rate of ≈ 25 cps to 275 kcps. The corresponding activity in the volume attributing to the count rate can be estimated to 1-5 kBq. On paper, this challenges the sensitivity of more established methods like BIXS (Beta-Induced X-ray Spectrometry) or even process ionisation-chambers developed at TLK while being more compact and/or less expensive, but this has to be demonstrated.

Having a closer look, the steps in Figure 2 nicely show the nearly instant response of the setup when changing the pressure of tritium at the GOS ceramic. However, it is noteworthy that, at least for a "fresh" setup that has not been operated for a longer time, a drift in the signal can be observed (insets to Figure 2) while the pressure reading remains constant within the sensitivity of the pressure transducer. Additionally, the always present memory effect cannot be neglected. Starting the first exposure to significant amounts of tritium at 275 kcps, the rate at approx. the same pressure after roughly 30 h of exposure is determined to be 630 kcps.

Figure 3 summarizes all measurements, consisting of a total of three measurement series with rising pressure followed by step-wise pressure decrease, as shown in Figure 2. With each cycling, the residual tritium level ("memory effect") increases, as can be seen at the elevated count rate after reaching the ultimate pressure ("0 mbar"). The rate ranges from 275 kcps for the newly connected setup to 2082 kcps after the 3rd exposure. Between those values, there was ≈ 4.5 bar·h of exposure to pure tritium. This offset cannot be neglected up to ≈ 20 mbar while for higher pressures nearly all values fall within a ± 5% error band around the exponential fit function. The handling of the memory effect is discussed in section 3.3.

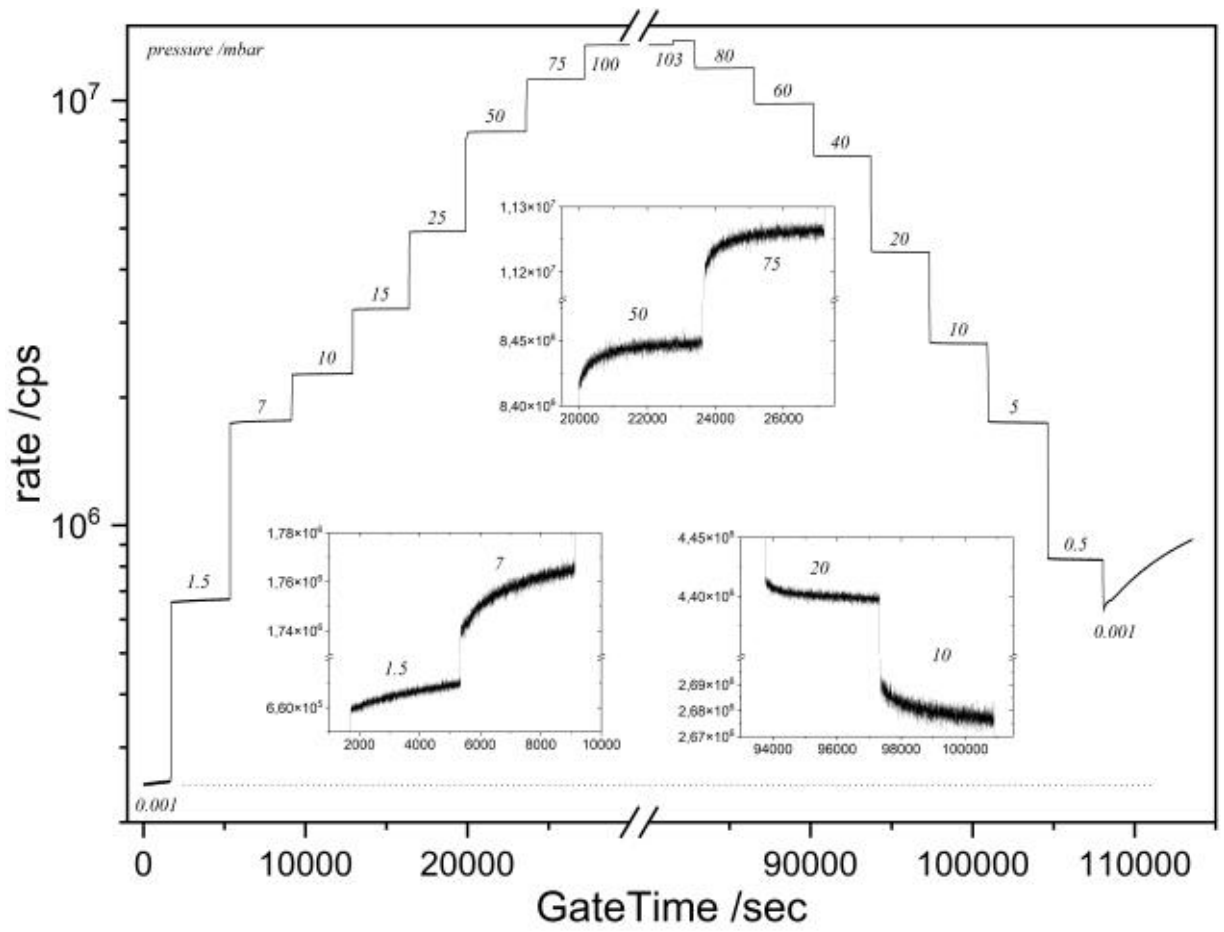

Figure 2: Stepwise increase and decrease of pressure of pure tritium and the setup's response. The figure on each step's plateau corresponds to the tritium pressure in the cell.

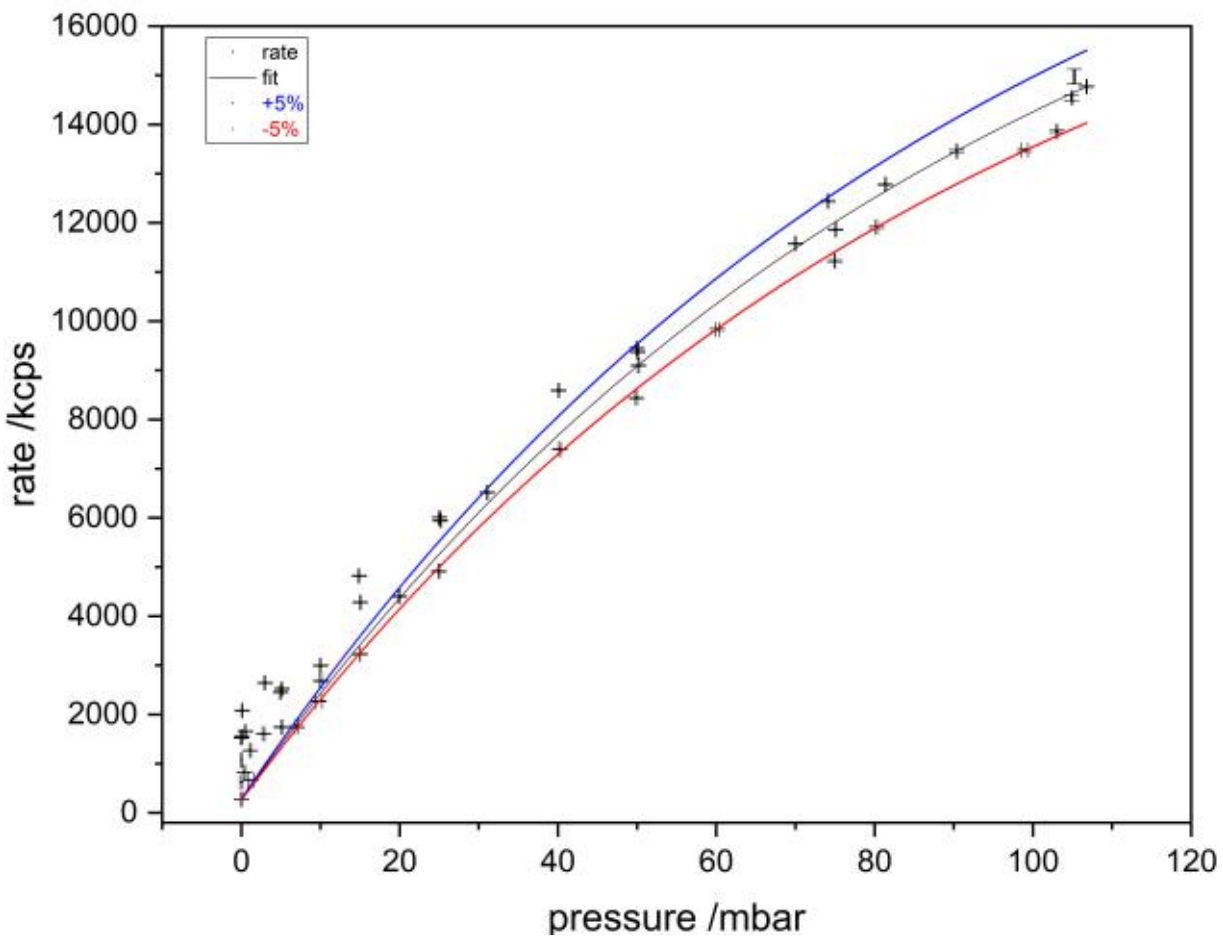

Figure 3: Combined results of three pressure increase/decrease cycles. The offset between the lowest and highest value in rate at "0 mbar" is attributed to "memory effect".

### 3.3 Memory effect and decontamination efforts

The observation of a memory effect was expected. Assuming all activity resides directly on top of the GOS ceramic, which could be considered a worst-case scenario, taking the mean activity, the conversion rate of the GOS, and geometry into account, a count rate of 2082 kcps translates to ≈ 19 kBq/cm². This figure is well in line with other experiments performed at TLK with metals (stainless steel, Cu, Au, Al) [10] and resides at the lower end in terms of contamination per unit time. Taking into account that not all activity is likely to be adhered to the GOS ceramic but rather distributed over all surfaces (stainless steel) in view of the ceramic, this figure is even lower.

However, if a memory effect is present, the lower level of detection (LoD) is affected negatively. Therefore, it is desirable to be able to reduce the memory effect.

The use of a flat, soldered, chemically inert ceramic opens up new possibilities in that regard. The sample volume can be easily dismantled and flushed with water, ethanol or acetone. Since water might affect the downstream vacuum system and acetone could damage parts of the glove box, we decided to use ethanol.

For this, the sample volume was thoroughly evacuated for ≈ 4 h, which reduced the rate to > 1600 kcps, disconnected and quickly filled with a few $cm^3$ ethanol. After 10 s, the ethanol was poured out and the cell re-connected. Immediately after, a rate of ≈ 475 kcps was registered which, after re-opening the valve connecting to the buffer vessel, steeply increased again (Figure 4) from residual tritium still present. A second decontamination with ethanol performed later reduced the rate further to ≈ 125 kcps (cell blind flanged to avoid back streaming of contamination from other parts of the setup), translating to roughly 1 kBq/cm² surface contamination.

## 4. Summary and conclusion

A new device based on scintillation for activity measurement of gaseous tritium has been presented. In contrast to other devices, this device uses ceramic scintillator material based on GOS:Tb with a high degree of compatibility to radioactive and aggressive media. Introducing a fully tritium and high vacuum compatible method for joining ceramic to stainless steel by a combination of sputter coating of various layers in combination with a soldering process, this is the first time the technique of scintillation counting for an inline gas monitor for large quantities of gaseous tritium is described. The system provides a low leak rate ($<10^{-10}$ mbar·l/s He equivalent), which is comparable with metallic sealings like Conflat® or VCR®, and complies fully with all requirements found at TLK or similar institutions.

To our knowledge, this is the first time a GOS scintillator has been successfully deployed to measure the low energy beta radiation of tritium and has handled a substantial amount of tritium during initial tests.

The selected detector system provides ease of use due to its single USB connection for both, power and data readout, and a very high sensitivity combined with a low dark-count background (25-50 cps).

The combination of both, vacuum tight soldered ceramic GOS scintillator and photomultiplier-system, enabled us to deploy a prototype of a high sensitivity, fully tritium compatible device. It provides a good usable range starting at < 1 kBq total activity and still performing well at > 10 GBq, as demonstrated in our tests with no degradation of the scintillator performance observed.

As always, material in contact with tritium shows a so-called memory effect, where a certain level of contamination remains on wetted surfaces. This is a limiting factor for all analytical equipment, independent of the detection principle or material selection. While "traditional" approaches using plastic scintillators can do little to mitigate this, GOS ceramics allow for the use of strong cleaning agents like ethanol (which would dissolve plastic scintillators) to effortlessly remove a large part (>95%) of the contamination within minutes without any specialised tools or processes.

## 5. Outlook

As described, the used photomultiplier module is very well suited for detection of low light levels, but struggles with high rates. This can be compensated by shortening the acquisition time from 1 sec to 0.1 sec but generally, this has its limits. The PMT module might not be necessary at all and could be replaced with a less expensive detector (e.g., photodiode) for higher partial pressures ("more" scintillation light). This would potentially enable the device to be operated in strong magnetic stray fields, too.

The influence of the thickness of the ceramics has to be investigated in order to find a good balance between photon yield and mechanical rigidity.

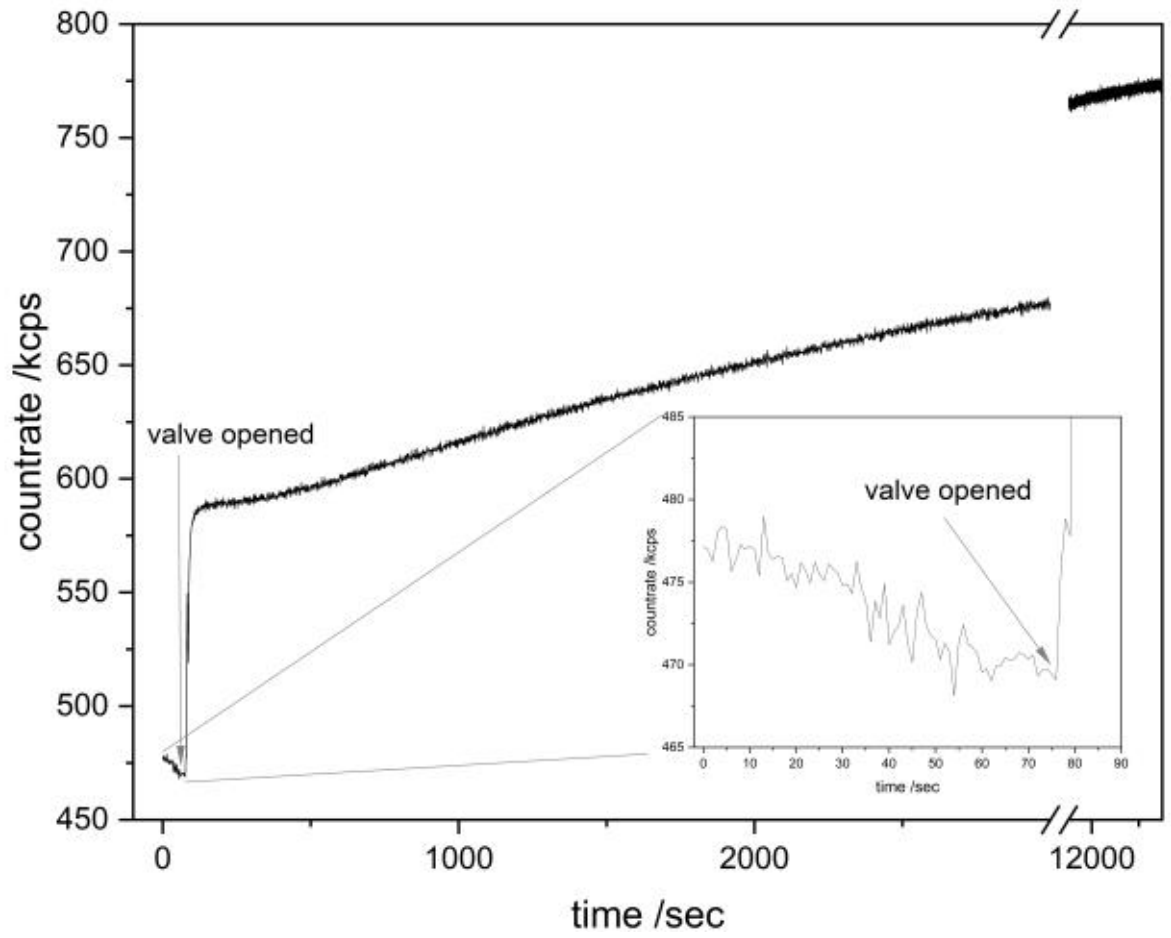

Figure 4: First decontamination with ethanol.

With regard to mixtures of tritium with other, inactive gases, any influence needs to be investigated since the measurements presented in this publication only used pure tritium. Self-absorption of electrons in the gas might start at a certain pressure of ad-mixed gas, as can be seen with BIXS devices. This would result in a lower-than-expected count rate for a certain tritium (partial-) pressure.

It is planned, to test the GaTAMoS setup against commercially available ionisation chambers. Clear advantages of the GaTAMoS setup are decontaminability, size, and price – the latter one especially if combined with a less expensive detector.

Another field of application is the measurement of tritiated water, where the amount of tritium contributing to a signal usually resides only in a very thin layer (few hundred µm) in direct contact with the scintillator due to the low penetration length of the low-energy β-electrons in the denser medium.